\documentclass[12pt,preprint]{aastex}

\newcommand{\be}{\begin{equation}}
\newcommand{\ee}{\end{equation}}
\newcommand{\mearth}{{ M_\earth }} 
\newcommand{\rearth}{{ R_\earth }} 
\newcommand{\rhat}{ {{\hat r}}} 
\newcommand{\vhat}{ {{\hat v}}} 
\newcommand{\xhat}{ {{\hat x}}} 
\newcommand{\yhat}{ {{\hat y}}} 
\newcommand{\cross}{ {{\Sigma_{\rm p}}}} 
\newcommand{\mrock}{ {{m_{\rm p}}} } 
\newcommand{\rrock}{ {{r_{\rm p}}} } 
\newcommand{\rhorock}{ {{\rho_{\rm p}}}} 
\newcommand{\fgp}{ f_{\rm p} }
\newcommand{\vpen}{ v_{\rm pen} } 
\newcommand{\vesc}{ v_{\rm esc} } 
\newcommand{\fact}{ {\cal F} } 
\newcommand{\sineint}{ {\rm Si} } 
\def\lta{\,\raise 0.3 ex\hbox{$ < $}\kern -0.75 em
 \lower 0.7 ex\hbox{$\sim$}\,}
\def\gta{\,\raise 0.3 ex\hbox{$ > $}\kern -0.75 em
 \lower 0.7 ex\hbox{$\sim$}\,} 

\begin{document} 

\title{\bf EFFECTS OF COLLISIONS WITH ROCKY PLANETS \\
ON THE PROPERTIES OF HOT JUPITERS}

\author{Kassandra R. Anderson$^{1}$ and Fred C. Adams$^{1,2}$}  

\affil{$^1$Physics Department, University of Michigan, Ann Arbor, MI 48109} 

\affil{$^2$Astronomy Department, University of Michigan, Ann Arbor, MI 48109} 

\begin{abstract}

Observed Hot Jupiters exhibit a wide range of physical properties.
For a given mass, many planets have inflated radii, while others are
surprisingly compact and may harbor large central cores. Motivated by
the observational sample, this paper considers possible effects from
collisions of smaller rocky planets with gas giant planets. In this
scenario, the Jovian planets migrate first and enter into $\sim4$ day
orbits, whereas rocky planets (mass $\mrock=0.1-20\mearth$) migrate
later and then encounter the gaseous giants. Previous work indicates
that the collision rates are high for such systems. This paper
calculates the trajectories of incoming rocky planets as they orbit
within the gaseous planets and are subjected to gravitational,
frictional, and tidal forces. These collisions always increase the
metallicity of the Jovian planets. If the incoming rocky bodies
survive tidal destruction and reach the central regions, they provide
a means of producing large planetary cores. Both the added metallicity
and larger cores act to decrease the radii of the gas giants at fixed
mass. The energy released during these collisions provides the Jovian
planet with an additional heat source; here we determine the radial
layers where kinetic energy of the colliding body is dissipated,
including the energy remaining upon impact with the existing core.
This process could have long-term effects if the colliding body
deposits significant energy deep in the interior, in regions of high
opacity. Both Hot Jupiters and newly formed gas giants have inflated
radii, large enough to allow incoming rocky planets to survive tidal
disruption, enhance the central core mass, and deposit significant
energy (in contrast, denser giant planets with the mass and radius of
Jupiter are expected to tidally destroy incoming rocky bodies).

\end{abstract}

\keywords{planets and satellites: composition --- planets and
  satellites: dynamical evolution and stability --- planets and
  satellites: formation}

\section{Introduction} 
\label{sec:intro} 

Since the first discovery of an exoplanet orbiting a main-sequence
star \citep{mayor}, the field of extrasolar planets has undergone
rapid progress. More than 750 planets have been discovered beyond our
Solar System, with over 1000 more possible objects awaiting further
measurements for confirmation \citep{borucki}.  Although the
observational sample is far from complete, it is clear that the Galaxy
hosts a diverse population of planets, with a range of masses spanning
over three orders of magnitude.  More than 150 of the $\sim750$
confirmed planets are gas giants with semi-major axes less than 0.1
AU. These planets are thought to reach such extreme environments after
forming further out in their circumstellar disks and migrating inward
\citep{migration}.  Up to $\sim 5\%$ of all solar systems are believed
to harbor ``Hot Jupiters'' \citep{bramich,gaudi}, so that a billion such
planets could reside in the Galaxy. 

The observed Hot Jupiters have a wide range of radii for a given mass
\citep{baraffe,burorton}. Many planets are larger than predicted, but
some are surprisingly small.  The reasons for these anomalous radii
remain uncertain, and the construction of a definitive mass-radius
theory for Hot Jupiters has become a central problem in exoplanet
research \citep[e.g.,][]{bodenheimer,laughlin}.  In general terms, the
inflated planets can be understood if they contain additional energy
sources \citep{bodenheimer,gu2004}, e.g., heat generated by Ohmic
dissipation \citep{batygin,perna,batygin2}, conversion of wind kinetic
energy into heat \citep{rauscher,guillot}, and/or tidal dissipation
\citep{socrates}.  In contrast, giant planets with deflated radii must
have high metallicities \citep{millerfort} and possibly large cores
\citep{burrows2}. The problem of these anomalous radii thus splits
into two sub-problems: [a] the source(s) of additional heating, and
[b] the ways in which planets can accumulate metallicity and large
cores.  Identifying possible long-term heat sources and examining the
ways in which planets can increase their core masses will help explain
the observed mass-radius relation and increase our understanding of
giant planet formation, structure, and evolution.

The goal of this paper is relatively modest: We consider the effects
of collisions of smaller rocky bodies with gaseous giant planets, and
explore how the collisions affect metallicity, core size, and perhaps
the energy budget. During the early epochs of solar system formation,
a large number of small terrestrial bodies are present; previous work
has shown that they can migrate inward and become captured by a Hot
Jupiter with high probability \citep{ketchum}. After capture, a rocky
planet orbiting within the interior of a Hot Jupiter loses energy due
to dissipative forces and eventually either collides with the core or
becomes tidally disrupted. These impacts provide a natural mechanism
for increasing the planetary metallicity, and could help explain some
of the inferred heavy cores and metal-rich compositions of the Hot
Jupiters. In addition, a substantial amount of energy is released
during these collisions. This energy could inflate a giant planet over
a longer period of time, provided that the energy is deposited deep
within the planet interior where the opacity is high.  On the other
hand, if the kinetic energy from the collision is dissipated in the
outer planetary layers, it would provide only a short-term source of
energy. These impacts thus have the potential to help explain, at
least in part, the observed radius anomalies.

Given the strong observational evidence for planets with large cores,
the process of core enhancement is of particular interest. A specific
observational example is provided by the planet HD 149026b, with a
total mass $M_P=0.36\pm0.03M_J$ and a radius $R_P=0.73\pm0.03R_J$
\citep{sato}. These observational constraints, along with theoretical
calculations \citep{fortney, ikoma}, indicate that this planet
contains $\sim80$ $\mearth$ of heavy elements. A significant fraction
of these heavy elements could reside in the core. Although current
observations do not fully constrain the core mass, it is expected that
HD 149026b and similar planets have larger cores than predicted by
theories of planet formation.  According to the core accretion model
\citep{pollack}, giant planet cores are formed from rocky
planetesimals in the circumstellar disk, and begin to accrete gas
after reaching a critical mass around $10\mearth$. Sufficiently large
planets experience runaway gas accretion and clear gaps in their
disks, so that the accretion of additional rocky material is
suppressed, making it difficult to build up core masses as large as
$80\mearth$. The planet HD 149026b, and others, motivates the
possible role of collisions in shaping giant planet cores.  A gaseous
planet with an initially small core mass could accrete one or more
rocky planets early in its lifetime and thereby accumulate a heavier
core.  Collisions with superearths ($\mrock\approx10\mearth$) are
probably the best way to explain the most massive cores, since the
accumulation of a large core from low-mass planets would require a
high encounter rate.  In addition, small planets are unlikely to
survive tidal forces over the entire trajectory and reach the core.
However, the accretion of smaller planets could affect the structure
of the giant planets by increasing the metallicity of the envelope.
 
In this paper we model the collisional dynamics between Jovian planets
and smaller rocky planets. This problem has been explored previously
in the context of Jupiter and Saturn \citep{anic,li} using smooth
particle hydrodynamical (SPH) simulations.  These authors have
investigated both the effects on the overall structure of the giant
planet and survival of the colliding bodies. The conditions required
for rocky planets to survive are particularly important. The previous
results indicate that low-mass planets ($\mrock\approx1\mearth$) are
often destroyed prior to reaching the core, whereas larger planets
($\mrock\approx10\mearth$) usually survive the entire trajectory. This
paper extends this previous work, but adopts a complementary approach.
SPH (or equivalent) methods provide a detailed description of the
collisions, but are computationally expensive and allow relatively few
cases to be considered. Here, instead, we treat the collision as a
one-body problem, where the incoming rocky body orbits within the
giant planet and is subject to gravitational, frictional, and tidal
forces. This approach allows for a much wider survey of parameter
space, which, in spite of the simplifying approximations, increases
our physical understanding of these collisions. More specifically,
this paper investigates the accretion of rocky bodies spanning two
orders of magnitude in mass ($\mrock=0.1-20\mearth$), within gas giant
planets of varying properties. Here we focus on Jovian planets with
mass $M_P=1M_J$ and radii in the range $R_P=1-2 R_J$, where the latter
are appropriate for young Hot Jupiters \citep{burrows}.

This paper is organized as follows: In Section \ref{sec:structure}, 
we construct polytropic models for gas giant planets both with and
without central cores. Using this model in Section \ref{sec:orbits},
we calculate the trajectories of incoming rocky bodies within the
gaseous giant planets. These results determine whether the rocky body
is captured or passes through the giant planet (as a function of
initial speed and impact parameter).  These trajectories also specify
the radial locations where energy is dissipated and the kinetic energy
remaining upon impact with the existing core. In Section
\ref{sec:tides}, we calculate the tidal forces exerted on an incoming
rocky planet by its host Jovian planet; such tidal effects are a
function of radial distance and determine whether or not the rocky
planet can survive. We conclude, in Section \ref{sec:conclude}, with a
summary of our results and a discussion of their implications.

\section{Planetary Structure Models}
\label{sec:structure} 

Before considering planetary collisions, we develop a structure model
for the Jovian (target) planet. Here we assume that the Jovian planet
consists of a rocky core of constant density surrounded by a gaseous
envelope.  The planet is assumed to be in hydrostatic equilibrium, and
obeys the equations of hydrostatic equilibrium and mass conservation, 
\be
\frac{dP}{dr} = -\frac{G M(r) \rho (r)}{r^2}
\qquad {\rm and} \qquad 
\frac{dM}{dr} = 4 \pi r^2 \rho (r) \, . 
\label{mass}
\ee
After specifying the equation of state $P(\rho)$, these equations can
be combined into a single differential equation, which can be solved
for the density profile. The conditions within Jovian planets require
complicated equations of state, which can vary from planet to planet. 
Since the goal of this paper is to develop a relatively simple model
to survey the parameter space for collisions, we adopt a polytropic
equation of state of the form
\be
P = K \rho^\Gamma \qquad {\rm where} \qquad 
\Gamma = 1 + {1 \over n} \, , 
\ee
and where $n$ is the polytropic index. Here we consider a range of
polytropic indices $n=1-2$, but detailed models suggest Jovian planets
are characterized by indices near the lower end ($n\sim1$) of this
range \citep[e.g.,][]{burtetons}. As shown below, the mean density of
the planet is one of the most important physical variables in the
problem; for polytropic models with fixed mass and radius, the mean
density does not differ greatly from that of more detailed models.
Further, the system to system variations in the structure of observed
exoplanets is larger than the differences resulting from this
polytropic approximation.

Following the formulation of \cite{chandrasekar}, we define 
dimensionless quantities $\xi$ and $f(\xi)$ through the relations 
\be
\xi = {r \over R} \quad {\rm and} \quad \rho = \rho_c f^n(\xi) \, , 
\qquad {\rm where} \qquad 
R^2 = {K \Gamma \over (\Gamma-1) 4\pi{G} {\rho_c}^{2 - \Gamma} } \, . 
\label{variables} 
\ee
With these definitions, the resulting differential equation for
$f(\xi)$ becomes the Lane-Emden equation
\be
\frac{1}{{\xi}^2}\frac{d}{d \xi} 
\left({\xi}^2 \frac{df}{d \xi} \right) + f^n = 0,
\label{LE}
\ee
with the boundary conditions
\be
f (\xi_c) = 1 \qquad {\rm and} \qquad 
\frac{df}{d\xi} \Bigg|_{\xi=\xi_c}= 0,
\ee
where $\xi_c$ denotes the radius of the planetary core. Planets
without central cores correspond to the limit $\xi_c \to 0$.  The
first zero of the function $f(\xi)$, denoted here at $\xi_0$, marks
the outer boundary (the surface) of the planet.

Although polytropic stellar structure models have been used
extensively \citep{chandrasekar,phillips}, this application differs
from most previous cases in two respects: First, we include the
possibility of a constant density core, so that the (dimensionless)
core radius $\xi_c$ provides an additional parameter. Second, the
expected values of the polytropic index are closer to $n$ = 1 for
giant planets, instead of the more familiar value $n$ = 3/2 for
degenerate stars and convective stars, or $n$ = 3 for massive stars
dominated by radiation pressure \citep{phillips}.

In terms of the dimensionless variables, the mass conservation 
equation is given by 
\be
\frac{d \mu}{d \xi} = {\xi}^2 f^n \, , 
\label{mu}
\ee
which is related to the physical enclosed mass profile $M(r)$ via
\be
M (r) = 4 \pi R^3 {\rho}_c \mu (\xi).
\label{dimmass}
\ee
The total dimensionless mass $\mu_0$ of the planet is thus
\be
{\mu}_0 = \int_0^{\xi_0} \xi^2 f^n(\xi) d\xi,
\ee
and the physical mass and radius are given by 
\be
M_P = 4 \pi R^3 {\rho}_c \mu_0 \qquad {\rm and} 
\qquad R_P = R\xi_0 .
\ee

Although the planetary structure is hydrostatic and does not require a
time variable, orbital motion is time dependent.  Here we define a
dimensionless time variable
\be
\tau = { t \over t_0} \qquad {\rm where} \qquad 
t_0 = \left( 4 \pi G {\rho}_c \right)^{-1/2} \, .  
\label{timedef} 
\ee
The time scale $t_0$ is the free-fall time appropriate for the density
at the planet center, and provides a benchmark dynamical time scale
for the orbit calculations. As an order of magnitude estimate, the
central densities of Jovian planets $\rho_c \sim 10$ g cm$^{-3}$, 
which corresponds to a time scale $t_0 \sim 345$ sec $\sim$ 6 min.

\begin{figure} 
\figurenum{1} 
{\centerline{\epsscale{0.90} \plotone{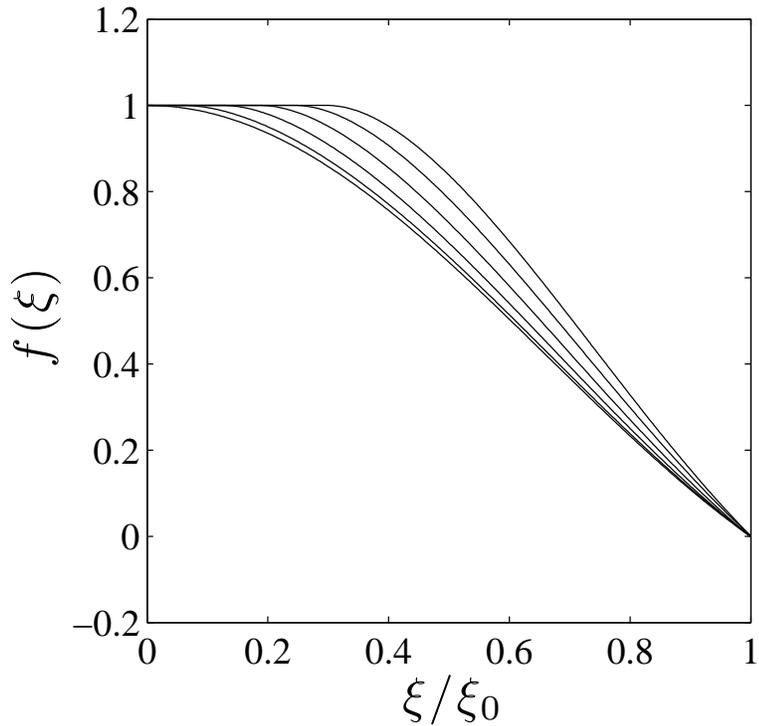} } } 
\figcaption{Density profiles for $n=1$ polytropic models with 
a collection of core masses. Each curve corresponds to a different
core radius from $\xi_0$ = 0 (bottom) to $\xi_c$ = 1 (top). For the 
largest value shown, $\xi_c$ = 1, the core makes up $\sim6.6\%$ of
the mass (e.g., the core would have mass $m_c\sim21\mearth$ for a
giant planet mass $M_P=1M_J$). }
\label{fig:polytrope}  
\end{figure}

Numerical integration of equation (\ref{LE}) yields a dimensionless
density profile, which can be converted to physical units by
specifying the central density.  The surface of the planet corresponds
to the radius $\xi_0$ where the density and pressure vanish.
Solutions for $f(\xi)$ are shown in Figure \ref{fig:polytrope} for a
polytropic index $n$ = 1 and dimensionless core radii $\xi_c=0-1$. 
Variation of the parameters $\xi_c$ and $n$ lead to a range of density
profiles, and upon specification of the physical planet mass and
radius, different central densities and pressures, core masses, and
core radii.  This model thus allows us to explore a variety of planet
structures. For this paper, we vary the core radius over a wide range,
with a focus on $\xi_c=0-1$, and vary the polytropic index over the
range $n=1-2$. For planets with Jovian masses and radii, we obtain
central densities $\rho_c=2-15$ g cm$^{-3}$, leading to central
pressures of order 100 Mbar, in agreement with more complicated models
\citep[e.g.,][]{militzer}. For a dimensionless core radius $\xi_c=1$, 
the core mass (in physical units) thus has values $m_c\approx 20-30$ 
$\mearth$ (for varying $n$). 

In general, equation (\ref{LE}) must be numerically integrated. For
the particular case where the polytropic index $n=1$, however, the
Lane-Emden equation (\ref{LE}) has an analytic solution of the form
\be
f(\xi) = A\frac{\cos{\xi}}{\xi} + B \frac{\sin{\xi}}{\xi}.
\label{gensolution}
\ee
For a coreless polytrope, the first term must vanish to avoid
divergence at the origin ($A = 0$), and the solution becomes 
\be
f(\xi) = \frac{\sin{\xi}}{\xi} \ . 
\ee
Note that the first zero occurs at $\xi_0 = \pi$. For finite core
radii ($\xi_c\ne0$), the general solution takes the form
\be
f(\xi)=\left(\xi_c\cos{\xi_c}-\sin{\xi_c}\right)\frac{\cos{\xi}}{\xi} 
+\left(\xi_c\sin{\xi_c}+\cos{\xi_c}\right) \frac{\sin{\xi}}{\xi}.
\ee
Note that for this general case, the first zero $\xi_0$ must be
determined by solving a transcendental equation. However, analytic
expressions for the mass profiles $\mu(\xi)$ can be found by
substituting these solutions for $f(\xi)$ into equation (\ref{mu}).
For the coreless polytrope, we find
\be
\mu(\xi) = - \xi \cos{\xi} + \sin{\xi} . 
\ee
For the general case, $\mu(\xi) = \xi^3/3$ for $\xi \leq \xi_c$, and 
\be
\mu(\xi) = \frac{1}{3}{\xi_c}^3 + (\xi_c - \xi) \cos{(\xi - \xi_c)} + 
(1 + \xi_c \xi) \sin{(\xi - \xi_c)}
\ee 
for $\xi > \xi_c$. Although the coreless polytrope represents a
relatively good model for planets with small cores like Jupiter, it is
less accurate for the exoplanets with massive cores.  Nonetheless, it
provides a reasonable approximation and the analytic forms for the
density and mass profiles are convenient for later calculations.

For a coreless polytrope of fixed index $n$, the model provides a two
parameter family of solutions, where the parameters are the central
density $\rho_c$ and the coefficient $K$ of the equation of state
(note that $K$ depends on the physics of the planetary interior and
can take a range of values). For any planet with observed mass $M_P$
and radius $R_P$, we can (in principle) adjust $\rho_c$ and $K$ to
provide the correct properties. This procedure results in a prediction
for the central density $\rho_c$ and the central pressure $P_c$ =
$K\rho_c^{\Gamma}$. As a consistency check, we fit the model to a
planet with the mass and radius of Jupiter, calculate the central
pressure and density, and then compare with current estimates. For
example, a coreless model with $n$ = 1 produces a central density
$\rho_c \approx 4$ g cm$^{-3}$ and a corresponding central pressure
pressure $P_c \approx 40$ Mbar. The actual central properties of
Jupiter are somewhat uncertain because the interior cannot be observed
directly and the conditions are difficult to recreate experimentally.
However, the central pressure of Jupiter is thought to lie in the
range $P_c$ = 10 -- 100 Mbar, with estimates around $P_c\approx70$
Mbar \citep{saumon}, in rough agreement with our predicted model
values. More detailed knowledge of these conditions requires a
reliable equation of state for hydrogen at high pressures, especially
in the transition to metallic hydrogen.  This poses significant
experimental and theoretical challenges, and many uncertainties remain
regarding the interiors of gas giants \citep{hubbard}.

The central pressure of the planet must obey the relation 
\be
P_c = \fact \frac{G M_P^2}{\pi R_P^4} = K \rho_c^\Gamma\, . 
\ee   
The first expression follows from the condition of hydrostatic
equilibrium (and dimensional analysis) whereas the second follows from
the polytropic equation of state. The dimensionless parameter $\fact$
is expected to be of order unity and can be written in the form
\be
\fact = \frac{\xi_0^4}{4 \mu_0^2 (n + 1)}.
\label{fpressure}
\ee  
For an $n$ = 1 polytrope with no core, $\xi_0 = \pi = \mu_0$, so that
$\fact$ = $\pi^2/8$. As the core size increases, the factor $\fact$
decreases, so that $\fact \approx 0.242$ for a core with $\xi_c$ =
2. For larger values of polytropic index $n$, the factor $\fact$ is
larger for planets with no core, but has nearly the same values for
planets with large central cores.

\section{Orbital Trajectories} 
\label{sec:orbits} 

In this section we solve the equations of motion for a rocky planet as
it falls into a gaseous giant planet. In this setting, we assume that
the Hot Jupiter has completed its inward migration and entered into a
stable orbit close to its star. Rocky planets migrate in later and
have a high probability of colliding with the Hot Jupiter, provided
that their orbital eccentricity is not overly damped \citep{ketchum}.
At the start of these integrations, the rocky planet already lies well
within the Hill sphere of the Jovian planet so that the gravitational
field of the star can be neglected. The system thus consists of a
Jovian planet with a mass $M_P = 1 M_J$ and a smaller rocky planet
with mass $\mrock \ll M_P$.  To first approximation, the system can be
treated as a one-body problem, with the origin of the coordinate
system placed at the center of the Jovian planet. As the rocky planet 
orbits within the gaseous envelope of the Jovian planet, it
experiences the usual gravitational force and a frictional (ram
pressure) force, so that the equation of motion is given by
\be 
\ddot{\bf{r}} =  - \frac{G M(r)}{r^2} \rhat 
- \frac{\rho(r) v^2 \cross}{\mrock} \vhat \, , 
\label{physequ}
\ee
where $\cross\sim{\pi}\rrock^2$ is the effective cross sectional area 
of the rocky planet (for determining ram pressure forces). Using
definitions from equation (\ref{variables}), we obtain 
\be 
{\ddot{\bf{r}} \over{R}} = {\ddot{\overrightarrow{\xi}}} = 
- {\mu(\xi)\over\xi^2} \rhat - \alpha f^n v^2 \vhat \, , 
\label{dimequ}
\ee
where $v$ is now the dimensionless speed (in units of $R/t_0$), and
where time derivatives are performed with respect to the dimensionless
time $\tau$ (from equation [\ref{timedef}]).  We have also defined a
dimensionless friction coefficient 
\be
\alpha \equiv {\xi_0^2 \over 4\pi\mu_0} 
\left({M_P \over \mrock}\right) 
\left( {\cross \over {R_{P}}^2} \right). 
\ee
Note that the coefficient $\alpha$ depends on the masses and radii of
both planets, as well as the polytropic index and the core radius.
Here we consider rocky planets with masses $\mrock=0.1-20\mearth$. For
a typical mean density $\langle \rho \rangle$ = 5.5 g cm$^{-3}$ for
the rocky planet, and for $M_P=M_J$ and $R_P=R_J$, the corresponding
values of the coefficient $\alpha$ lie in the range
$\alpha\approx0.5-5$. Note that Jovian planets with inflated radii
have lower values of $\alpha$ and provide less friction.

The relevant parameter space thus consists of five variables: the
impact parameter $x_0$ of the collision, initial velocity $v_0$,
friction coefficient $\alpha$, the polytropic index $n$, and the
dimensionless core radius $\xi_c$. Once these parameters are
specified, the equation of motion (\ref{dimequ}) can be numerically
integrated. 

\subsection{Rocky Planets Orbiting within Jovian Planets} 
\label{sec:orbitshapes} 

To begin, we consider the simplest case of a coreless Jovian planet
($\xi_c = 0$) and adopt a polytropic index $n=1$.  By fixing the 
polytropic index and core radius, the number of free parameters is
reduced to three. The starting conditions are defined as follows: The
calculations are started with the incoming rocky planets located at
the outer edge of the Jovian planet, traveling in the $-\yhat$
direction. In dimensionless coordinates, the rocky planet has initial
position $(x_0,y_0)$, where $x_0$ is identified as the impact
parameter, so that the starting coordinates obey the constraint
\be
x_0^2 + y_0^2 = \xi_0^2 \, . 
\label{constraint}
\ee
Here we explore the full range of impact parameters
$0\le{x_0}\le\xi_0$. For the sake of definiteness, we consider
incoming rocky planets with masses $\mrock=1$ and 10 $\mearth$.  If
the Jovian planet has the mass and radius of Jupiter, these values
correspond to dimensionless friction coefficients in the range
$\alpha$ = 1 -- 3.  Dynamical simulations \citep{ketchum} indicate
that the expected impact velocities lie in the range $v_0=40-150$ km
s$^{-1}$.  This study considers a wider range of impact velocities
where $v_0=1-10^4$ km s$^{-1}$.

\begin{figure} 
\figurenum{2} 
{\centerline{\epsscale{0.90} \plotone{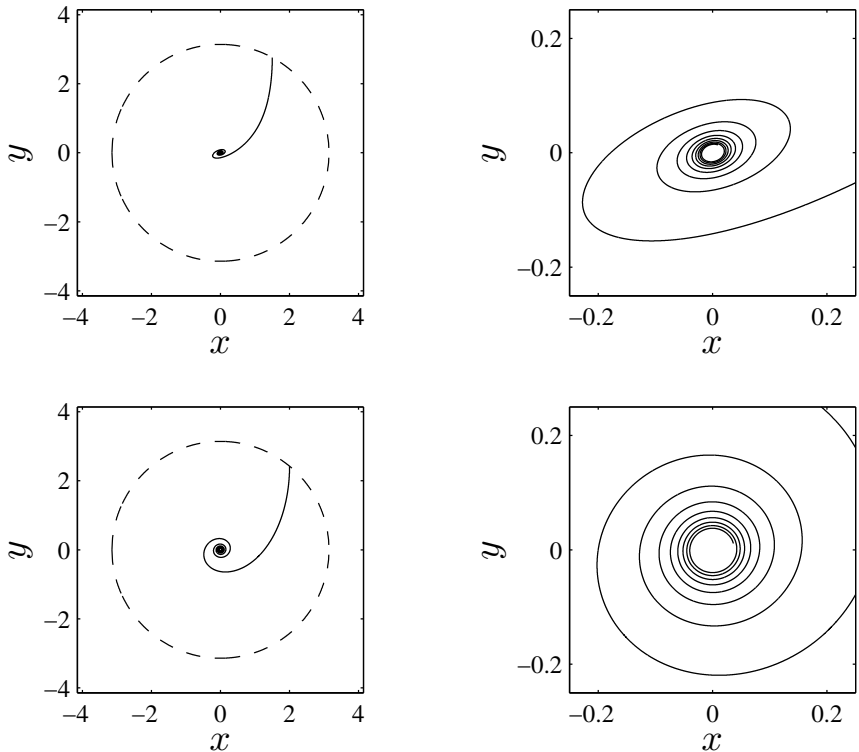} } } 
\figcaption{Orbits within coreless planets with polytropic index 
$n$ = 1. The top panels show results for incoming rocky planets 
with masses $\mrock=1\mearth$; and bottom panels show results for 
$\mrock=10\mearth$. The panels on the left show a full planetary view,
whereas the panels on the right show the orbits on a smaller spatial
scale near the center. Both collisions use an impact parameter $x_0$ =
1.5 and an impact velocity $v_0$ = 40 km s$^{-1}$. }
\label{fig:orbits} 
\end{figure}

Examples of typical trajectories are shown in Figure \ref{fig:orbits}
for a dimensionless impact parameter $x_0$ = 1.5, an initial velocity
$v_0$ = 40 km s$^{-1}$, and rocky planet masses $\mrock$ = 1 and 10
$\mearth$ (friction parameters $\alpha=1-3$).  As the rocky planets
spiral inward within the Jovian planets, the orbits of larger rocky
planets remain relatively circular, while the orbits of smaller rocky
planets become highly elliptical.  In general, the bodies continuously
spiral inward and approach the origin asymptotically.

\begin{figure} 
\figurenum{3} 
{\centerline{\epsscale{0.90} \plotone{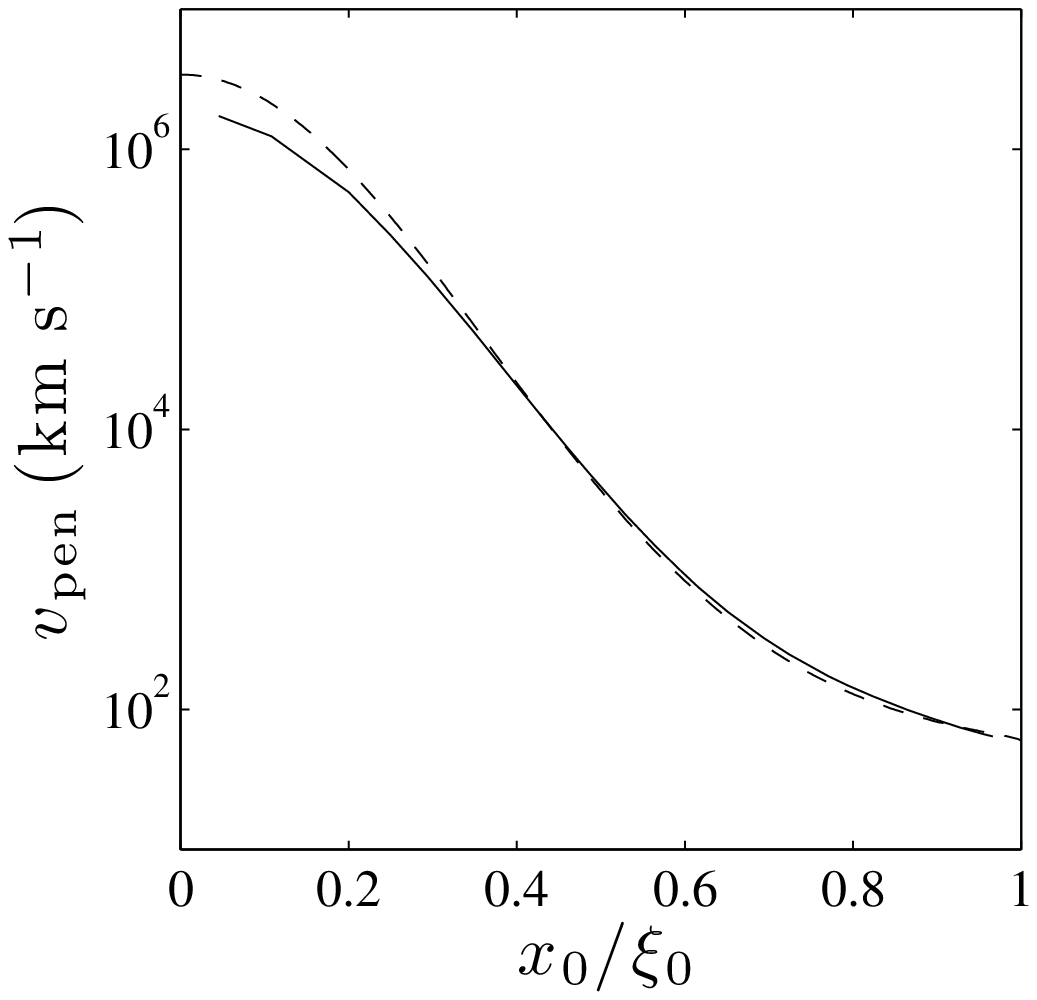} } } 
\figcaption{Penetration velocity as a function of impact parameter.
The velocity $\vpen$ is the impact velocity (in km s$^{-1}$) needed 
for the rocky planet to pass through the gaseous giant planet, emerge 
from the far side, and remain uncaptured. The impact parameter is
given as a fraction of the planetary radius. Here the giant planet has
mass $1M_J$, radius $1R_J$, and $n$ = 3/2.  The solid curve shows the
numerical results; the dashed curve shows the analytic approximation
derived in Appendix \ref{sec:speedappendix}. The approximation works
well over the full range of impact parameters, with the largest
discrepancy for head-on collisions $(x_0=0)$. }
\label{fig:vpen} 
\end{figure}

There are two possible outcomes of these orbits: Either the rocky
planet is captured and spirals inward, or it passes through the
gaseous planet and retains enough kinetic energy to escape. For an
incoming rocky planet with given impact parameter $x_0$, we can
determine the initial velocity necessary for the rocky planet to avoid
capture. More specifically, the critical condition for penetration
requires that the rocky planet initially has enough kinetic energy so
that it passes through the gaseous planet and leaves its surface
traveling outward at the escape speed.  This penetration speed is
denoted here as $\vpen$ and is a function of $x_0$. For a given impact
speed, incoming rocky planets will be captured for all impact
parameters $x_0 < x_\ast$, where the maximum impact parameter defines
the effective capture cross section $\sigma(v_0)$ = $\pi{x}_\ast^2$. 

Here we calculate the required penetration speeds numerically using
the above formulation. We also derive an analytic estimate of the
penetration speed $\vpen$ in Appendix \ref{sec:speedappendix}. Figure
\ref{fig:vpen} shows the numerically determined penetration speeds
along with the analytic approximation. The best agreement between the
numerical and analytic results occurs for large impact parameters
because the assumption of constant density (see Appendix
\ref{sec:speedappendix}) is more accurate far from the center of the
planet; even for small impact parameters, however, the results agree
within a factor of two. These calculations show that enormous starting
velocities are required for penetration when the impact parameter is
near zero, i.e., $\vpen\sim10^6$ km s$^{-1}$. For somewhat larger
impact parameters $x_0\sim\xi_0/2$, penetration still requires
$\vpen\sim1000$ km s$^{-1}$. As a result, the cross section for
capture is large. For the expected impact speeds $v_0=40-150$ km
s$^{-1}$, incoming rocky bodies will penetrate the Jovian planet only
for impact parameters comparable to the radius of the planet, roughly
$x_0\gta0.8\xi_0=x_\ast$. For high-speed rocky planets with $v_0>100$
km s$^{-1}$, the effective target area (capture cross section) of the
Jovian planet is thus reduced by $\sim30\%$.

\begin{figure} 
\figurenum{4} 
{\centerline{\epsscale{0.90} \plotone{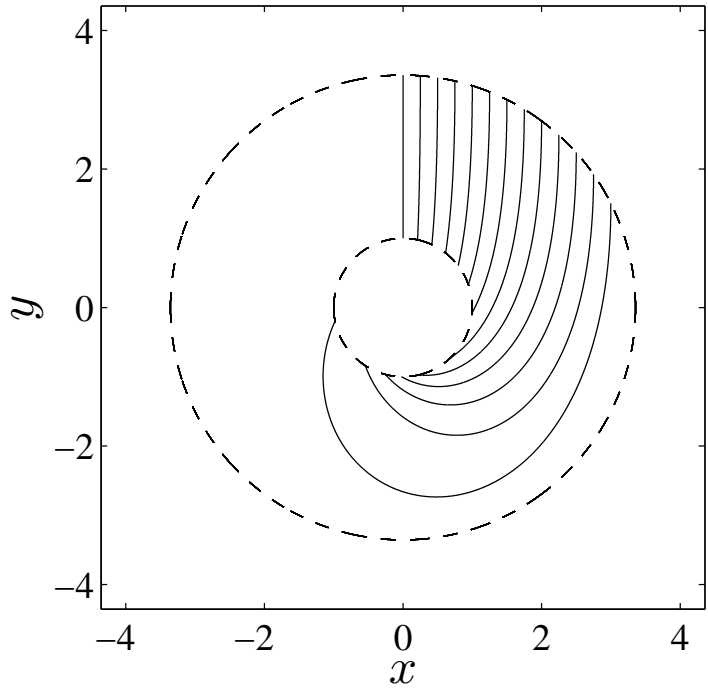} } } 
\figcaption{Trajectories for different impact parameters. Each curve 
shows the orbit of an incoming rocky body for a given impact parameter. 
The outer dashed circle marks the location of the outer surface of the 
Jovian planet; the inner dashed circle marks the location of its core, 
which has dimensionless radius $\xi_c$ = 1. The impact speeds, defined 
at the surface of the Jovian planet, are taken to be 80 km s$^{-1}$. } 
\label{fig:traject} 
\end{figure}

We now consider the general case where rocky planets collide with
Jovian planets containing central cores.  For the sake of
definiteness, we begin by choosing a standard (dimensionless) core
radius $\xi_c$ = 1.0.  Depending on the polytropic index, this core
radius spans $\sim20-30\%$ of the giant planet radius, and encloses a
mass of $20-30\mearth$.  Here we use three polytropic indices ($n$ =
1, 3/2, and 2), four terrestrial planet masses ($\mrock$ = 0.1, 1, 10,
and 20 $\mearth$), and the full range of impact parameters
$x_0=0-\xi_c$, where the initial velocities span the range $v_0$ = 20
-- 150 km s$^{-1}$.  The resulting trajectories, including profiles of
energy dissipation and the velocities at the core, are nearly
independent of the polytropic index.  Given that similar results were
obtained for all three choices ($n$ = 1, 3/2, and 2), only the results
for the $n = 1$ polytrope are shown here.  Figure \ref{fig:traject} shows
the trajectories for rocky planets with initial speeds of 80 km s$^{-1}$ 
and with varying impact parameter.  

We note that the presence of a core has little effect on whether or
not an incoming rocky planet will penetrate the Jovian planet and
escape capture. As a result, the results depicted in Figure
\ref{fig:vpen} remain valid --- only incoming planets with large
impact parameters ($x_0 > 0.8 \xi_0$) or incredibly large initial
speeds ($v_0 > 1000 $ km s$^{-1}$) are able to escape.

\subsection{Energy Dissipation}
\label{sec:dissipation} 

\begin{figure} 
\figurenum{5} 
{\centerline{\epsscale{0.90} \plotone{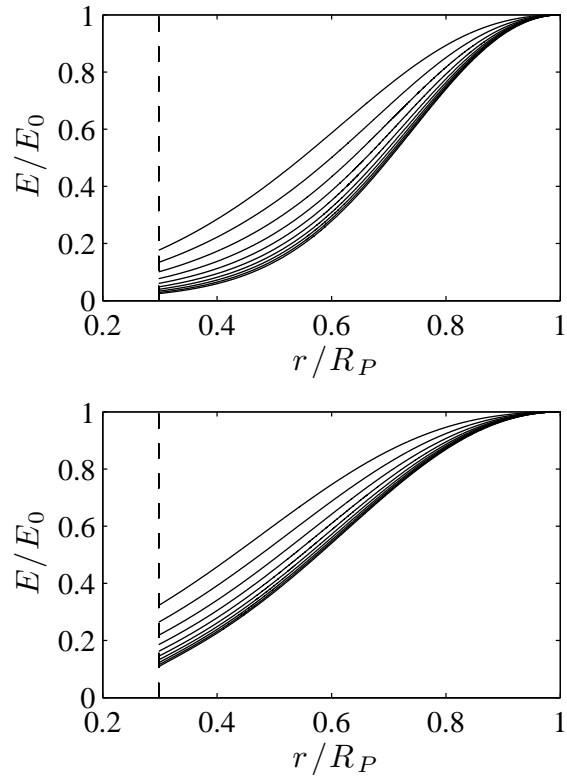} } } 
\figcaption{Fraction of the total energy of the rocky planet as a 
function of radial distance for varying initial velocities. Masses for
the rocky planet are $\mrock$ = 1 $\mearth$ (top panel) and $\mrock =
10 \mearth$ (bottom panel). The Jovian planet has polytropic index $n$
= 1, mass $M_P=1M_J$, and radius $R_P=1R_J$. The dashed lines indicate
the core radius. }
\label{fig:energyv0} 
\end{figure}

\begin{figure} 
\figurenum{6} 
{\centerline{\epsscale{0.90} \plotone{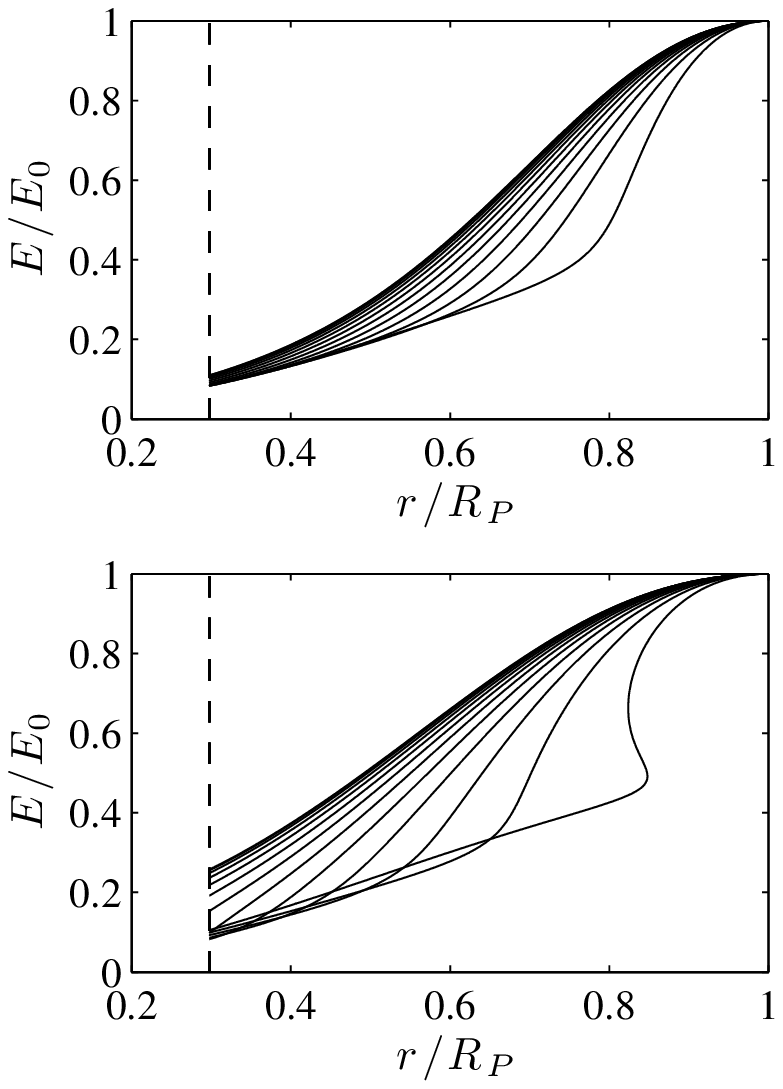} } } 
\figcaption{Fraction of the total energy of the rocky planet as a 
function of radial distance for varying impact parameters. Masses for
the rocky planet are $\mrock$ = 1 $\mearth$ (top panel) and $\mrock =
10 \mearth$ (bottom panel). The Jovian planet has polytropic index $n$
= 1, mass $M_P=1M_J$, and radius $R_P=1R_J$. The dashed lines indicate
the core radius. }
\label{fig:energyx0} 
\end{figure}

During each incoming trajectory, energy is dissipated through the
action of frictional forces. Figures \ref{fig:energyv0} and
\ref{fig:energyx0} show the total energy (kinetic and potential) of
the rocky planet as a function of radial distance, measured from the
center of the giant planet. These figures show results for target giant
planets with the mass and radius of Jupiter, and for incoming rocky
planets with $\mrock=1\mearth$ (top panels) and $\mrock=10\mearth$
(bottom panels). Figure \ref{fig:energyv0} shows energy profiles for
varying initial impact speeds, whereas Figure \ref{fig:energyx0} shows
energy profiles for varying impact parameters. In all cases shown, the
majority of the energy is dissipated in the outer layers ($r/R_P>0.5$)
of the planet. 

As shown in Figure \ref{fig:energyx0}, for sufficiently large impact
parameters the energy fraction $E/E_0$ takes on multiple values for a
given radius; on the other hand, the radius is a single-valued
function of energy. This complication arises for orbits that reach an
inner turning point and reverse course before finally spiraling inward
(e.g., see the orbit with the largest impact parameter in Figure
\ref{fig:traject}). For rocky planets with larger masses and/or giant
(target) planets with lower density, the orbits display additional
turning points and the energy dissipation curves show even more
structure (these cases are not shown in the figures). 

When the rocky planet strikes the core, its motion is arrested and any
remaining kinetic energy is converted into heat. The energy remaining
upon impact with the core thus determines the extent of possible core
deformation as well as the amount of energy available for long-term
heating. This fraction of the initial energy remaining depends on the
mass of the rocky planet and the structure of the Jovian planet.
Rocky planets with larger masses retain a larger fraction of their
initial energy (as well as a larger absolute value of energy). For
target giant planets with the properties of Jupiter (see Figures
\ref{fig:energyv0} and \ref{fig:energyx0}), 1 $\mearth$ planets lose
most of their initial energy, whereas 10 $\mearth$ planets retain up
to $30\%$.  The structure of the Jovian planet, in particular its mean
density, plays the most important role. When the radius of the giant
planet is doubled (which may be appropriate for young planets), the 10
$\mearth$ planets can retain the majority of their initial energy (not
shown).  An analytic desription of the energy dissipation process is
given in Appendix \ref{sec:energyappendix}.

\begin{figure} 
\figurenum{7} 
{\centerline{\epsscale{0.90} \plotone{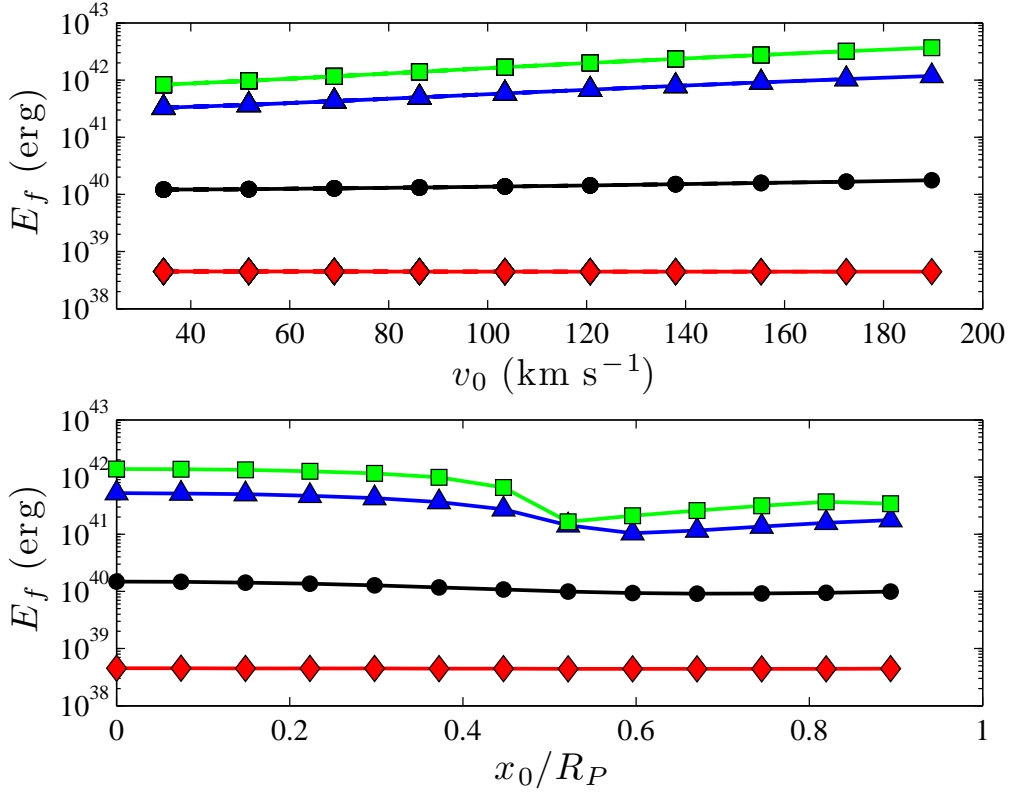} } } 
\figcaption{Kinetic energy of the rocky planet as it reaches the core 
of the Jovian planet with radius $R_P=R_J$. Impact energy is shown as
a function of initial velocity for fixed $x_0$ = 1 (top panel) and as
a function of impact parameter at fixed impact speed $v_0$ = 80 km
s$^{-1}$ (bottom panel). The four curves correspond to four masses for
the rocky planet, for $\mrock$ = 0.1, 1, 10, and 20 $\mearth$ (from
bottom to top).  The Jovian planet has polytropic index $n$ = 1, mass
$M_P=M_J$, and dimensionless core radius $\xi_c$ = 1. Note that the
curves do not extend beyond $x_0 \approx 0.9$; for collisions with
larger impact parameters, rocky planets pass through the gaseous
planet and escape. }
\label{fig:impact} 
\end{figure}

The kinetic energy of the rocky planet at the moment of impact with
the core is shown in Figure \ref{fig:impact} as a function of both the
initial velocity and impact parameter. The impact velocities vary over
the range $10-75$ km s$^{-1}$, leading to impact energies in the range
$10^{38}-10^{42}$ ergs, where these values are strongly dependent upon
the rocky planet mass.  Planets less massive than $\mrock=1\mearth$
show little spread in their final energy (velocity) for the entire set
of impact parameters and initial velocities. This result can be
understood in terms of the nature of the frictional force
($F_f\sim\rho v^2$) and the size of the friction coefficient $\alpha$.
Planets with the highest velocities experience the highest levels of
damping, and their speeds are quickly diminished.  Furthermore, the
impact parameter determines the average density encountered by the
orbiting planets.  Although the average density is lower for planets
incident at large impact parameters, these planets must travel farther
before being accreted onto the core, and are thus subjected to damping
forces acting over larger distances. These effects are the most
pronounced for low-mass planets, since they experience the greatest
frictional forces relative to gravitational forces.  Planets with
masses larger than $\mrock\sim1\mearth$ are subjected to relatively
less friction and are more sensitive to variations in the initial
velocity and impact parameter.  Planets incident at small impact
parameters (roughly, less than the core radius) collide directly with
the core and travel shorter distances.  These planets thus experience
less damping and retain a significant amount of energy upon impact
with the core.  Incoming planets with sufficiently large impact
parameters do not collide directly with the core, but rather gradually
spiral inward and often enter into highly elliptical orbits. The
orbital path length thus increases substantially for impact parameters
larger than a critical value, and the amount of energy deposited at
the core decreases accordingly. For the regime of parameter space
considered here, this critical impact parameter is roughly
$x_{0\ast}\approx 0.4 R_P$. Finally, we note that the curves in 
Figure \ref{fig:impact} do not extend beyond $x_0 \approx 0.9 R_P$; 
for collisions with larger impact parameter, the rocky planet does 
not remain bound and hence does not reach the core. 

\begin{figure} 
\figurenum{8} 
{\centerline{\epsscale{0.90} \plotone{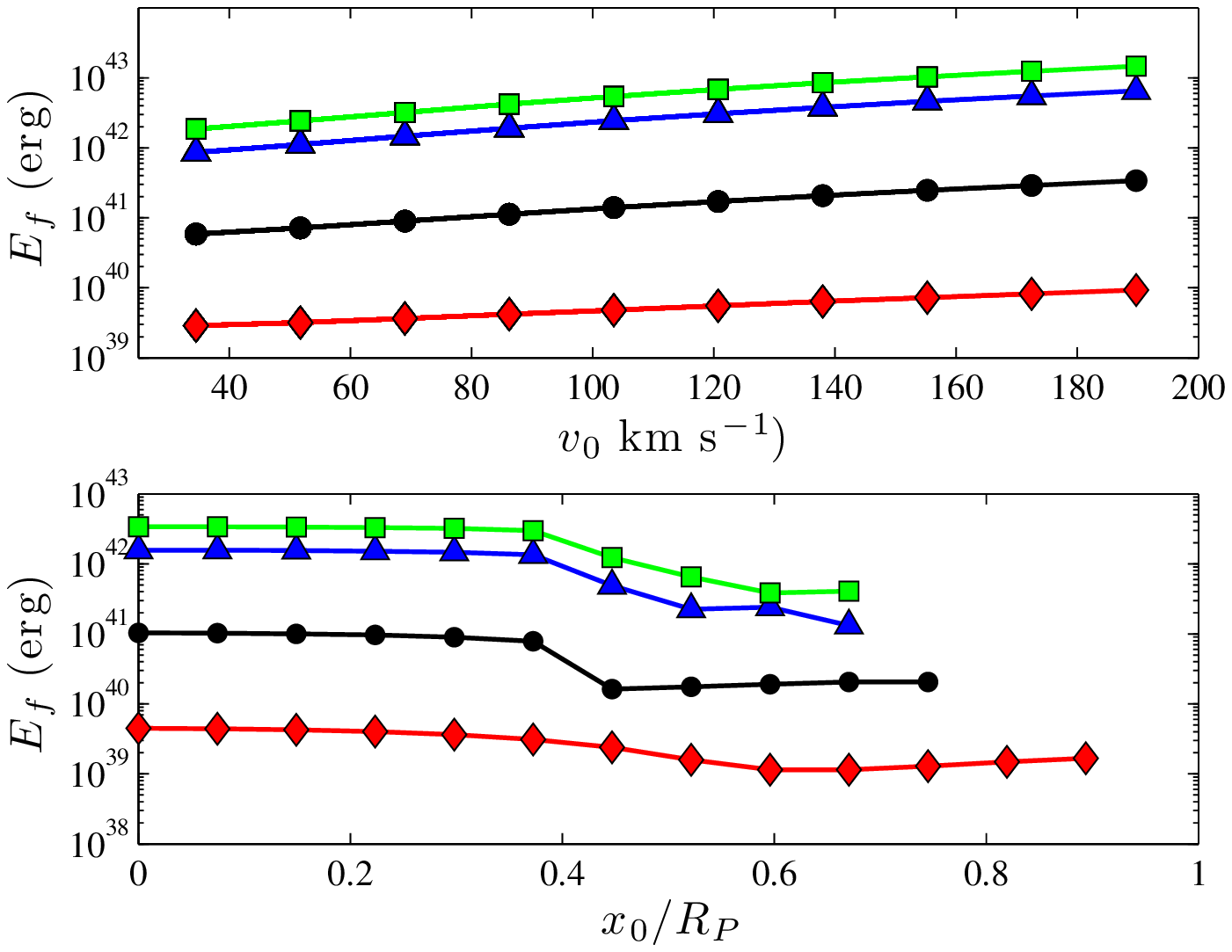} } } 
\figcaption{Kinetic energy of the rocky planet as it reaches the core 
of a young Jovian planet with large radius $R_P=2R_J$. Impact energy
is shown as a function of initial velocity for fixed $x_0$ = 1 (top
panel) and as a function of impact parameter at fixed impact speed
$v_0$ = 80 km s$^{-1}$ (bottom panel).  The four curves correspond to
four masses for the rocky planet, for $\mrock$ = 0.1, 1, 10, and 20
$\mearth$ (from bottom to top).  The Jovian planet has polytropic
index $n$ = 1, mass $M_P=M_J$, and dimensionless core radius $\xi_c$ =
1. Note that the curves do not extend all the way to $x_0$ = 1; for
collisions with sufficiently large impact parameters, rocky planets
pass through the gaseous planet and escape. }
\label{fig:impact2} 
\end{figure}

In the calculations presented so far, the mass and radius of the
Jovian planet were taken to be those of Jupiter.  However, many Hot
Jupiters have inflated radii, with typical values $R_P \sim 1.2 R_J$
for a mass $M_P$ = 1 $M_J$.  In addition, the radii of young giant
planets --- before they contract and cool --- are even larger than
those of mature planets \citep{burrows}. As a result, collisions of
rocky planets with inflated gaseous planets (for the same mass) must
be explored. To bracket the possibilities, the previous calculations
were repeated for a Jovian planet with mass $M_P=1M_J$, but with a
larger radius $R_P=2R_J$, leading to a lower mean density of only
$\rho\sim0.1$ g cm$^{-3}$, ten times lower than the mean density of
Jupiter. These results are shown in Figure \ref{fig:impact2}. Compared
to the previous case (with $R_P=1R_J$), the impact energies are larger
by an order of magnitude. Notice also that the curves do not extend to 
large impact parameters (bottom panel) because the rocky planets pass 
through the gaseous planet. 

\begin{figure}
\figurenum{9} 
{\centerline{\epsscale{0.90} \plotone{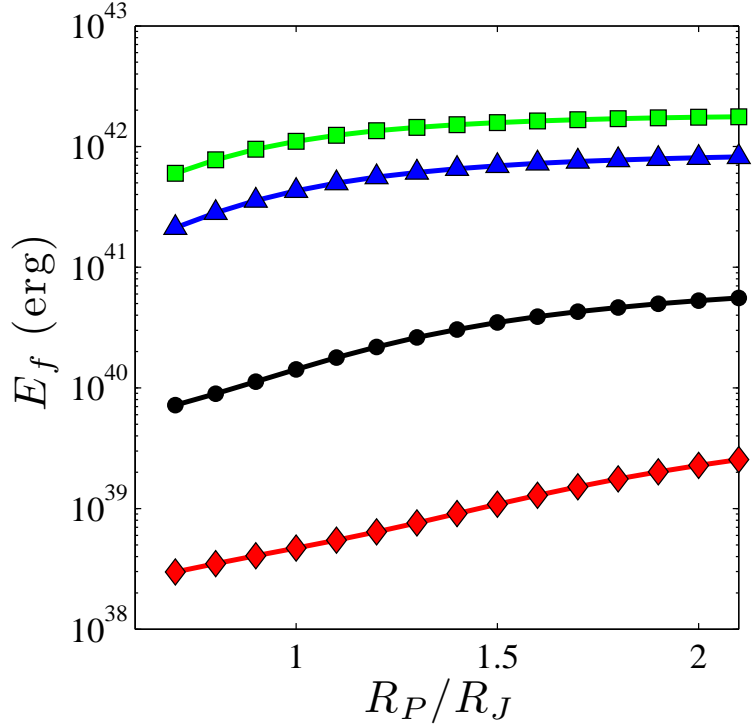} } } 
\figcaption{Effects of the Jovian planet radius on collisions with 
rocky planets. Kinetic energy of the rocky planet as it reaches the
core of the Jovian planet is shown as a function of Jovian planetary
radius $R_P$. The four curves correspond to rocky planet masses
$\mrock$ = 0.1, 1, 10, and 20 $\mearth$ (from bottom to top). The
Jovian planet has polytropic index $n$ = 1, mass $M_P=M_J$, and
dimensionless core radius $\xi_c$ = 1. For all cases shown, the
initial speed (evaluated at the surface of the Jovian planet) is 
$v_0$ = 50 km s$^{-1}$ and the impact parameter $x_0$ = 0 
(head-on collisions). } 
\label{fig:impactrad} 
\end{figure}

Another way to illustrate the effects of inflated radii for the Jovian
planets is to the plot the impact energy as a function of the radius
$R_P$ of the gas giant. These results are shown in Figure
\ref{fig:impactrad} for collisions with rocky planets that have
starting speeds $v_0$ = 50 km s$^{-1}$ and varying masses $\mrock$ =
0.1, 1, 10, and 20 $\mearth$. The kinetic energy $E_f$ remaining when
the rocky planet strikes the core of the Jovian planet is a smoothly
increasing function of the radius of the larger body. To leading
order, the four curves shown in Figure \ref{fig:impactrad} have nearly
the same shape, so that they are scaled by the mass of the rocky
planet. The starting energy for the 1 $\mearth$ planet is 7.5 $\times
10^{40}$ erg; for collisions with Jovian planets with $R_P = 2.1 R_J$,
the impact energy is about two thirds of this starting value. Rocky
planets with smaller masses lose a somewhat greater fraction of their
initial energy, whereas sufficiently massive rocky planets ($\mrock
\gta 10 \mearth$) gain more energy by falling through the
gravitational potential well of the giant planet than they lose
through frictional dissipation.

This difference in the energy dissipation between Jovian planets with
different radii is remarkable.  The planets with inflated radii tend
to dissipate far less energy in their outer envelopes, and the
percentage of the initial energy remaining when the incoming rocky
planets reach the core can be as high as $70\%$.  Note that this
percentage represents the fraction of the rocky planet's total energy
(kinetic and potential) at the surface of the Jovian planet.  If we
examine purely the kinetic energy, we find that the energy dissipation
in the outer regions is low enough that the kinetic energy can
actually increase during the inward trajectory, as it travels deeper
into the potential well of the giant planet.

The consequences of these collisions differ greatly for young Jovian
planets with low densities and mature planets with higher densities.
Mature planets are able to dissipate much more of the incoming energy
near the surface, where heat is transported efficiently by radiation.
In such cases, the energy of the impact is expected to radiate away
quickly, and the structure of the giant planet should recover on short
time scales. On the other hand, young planets with large radii tend to
dissipate little energy in the radiative zones, and the majority of
the impact energy is delivered to the core, where it becomes trapped
and is available for long-term heating.  The evolutionary stage of the
Jovian planet is therefore an important factor in determining the
long-term effects of these collisions. Young inflated planets will be
far more susceptible to core enhancement and prolonged heating than
mature planets. Similarly, Hot Jupiters will be more susceptible than
giant planets farther from their stars.

\section{Tidal Disruption}
\label{sec:tides} 

For purposes of finding their orbits, the impinging rocky planets have
thus far been treated as indestructible point particles. In this
section we consider the possibility that these planets can be
disrupted by tidal forces. If the incoming planet is destroyed, several
outcomes are possible.  Depending on the interior conditions of the
giant planet, the rocky planet remnants could either become accreted
onto the core, or uniformly enrich the planetary interior with heavy
elements. A full determination of the outcome is computationally
expensive and is beyond the scope of this present work; in addition,
it requires a greater understanding of Jovian planet interiors (e.g.,
including convection) than is currently available.  In this section we
consider how tidal forces act on the incoming rocky planets and
determine the likelihood of these bodies reaching the core (of the
Jovian planet) relatively intact. These results allow us to estimate
the maximum amount of mass and energy deposited at the core.

At any point in the trajectory, the tidal force (per unit mass)
exerted on the rocky planet can be approximated by the difference in
the gravitational acceleration at the near and far surfaces,
\be
\Delta F = \frac{GM(r)}{r^2} - 
\frac{GM(r + \Delta r)}{(r + \Delta r)^2},
\label{tidalforce} 
\ee  
where $\Delta{r}=2\rrock$ (the rocky planet diameter). Note that the
tidal forces arise due to the difference in the enclosed mass as well
as the difference in the radial distance.  Since the tidal forces are
not spherically symmetric with respect to the rocky planet, this
treatment introduces an approximation.  Nonetheless, if the rocky
planet is to withstand tidal forces, the tidal acceleration given by
equation (\ref{tidalforce}) must be less than the surface gravity of
the rocky planet.  Equating these quantities yields the critical
condition for survival, i.e.,
\be
\left| \frac{GM(r)}{r^2} - \frac{GM(r + \Delta r)}{(r + \Delta r)^2} 
\right| = \frac{G \mrock}{{\rrock}^2} \, .
\label{tidalexact}
\ee   
This expression can be simplified by expanding in $(\Delta r)/r\ll{r}$ 
and keeping only the leading order terms. With these approximations,
the expression (\ref{tidalforce}) for the tidal force takes the more 
familiar form
\be
\Delta F = \frac{4 \rrock G M(r)}{r^3} - 8 \pi \rrock G \rho(r).
\label{usualtide} 
\ee
Near the surface of the Jovian planet, the density is small enough so
that the second term on the right hand side is small compared to the
first.  After dividing by $\rrock$, we obtain a rough criterion for
the survival of the rocky planet, 
\be
\frac{4 M(r)}{r^3} \le \frac{\mrock}{{\rrock}^3}.
\label{tidalapprox}
\ee 
In other words, the incoming body will be tidally disrupted if its
mean density is less than (about) four times the mean density of the
giant planet.  Note that equation (\ref{tidalapprox}) is valid near
the surface of the Jovian planet, where the differential change in
mass and the ratio $(\Delta r)/r$ are relatively small. Here we
consider rocky planets with radii $\rrock\sim 1\rearth$, and giant
planets with radii $R_P\sim10\rearth$, so that $(\Delta{r})/r$
$\approx0.2$ at the surface and increases with decreasing radial
position. As a result, the approximation scheme deteriorates as the
rocky planet moves inward. Here we use equation (\ref{tidalapprox}) to
provide an order of magnitude estimate for the tidal stretching at the
Hot Jupiter surface, but use the full form of equation (\ref{tidalforce}) 
to assess the prospects for rocky planet survival (see below). 

Notice that the two sides of equation (\ref{tidalapprox}) are nearly
equal for a pair of planets with the mean densities of Jupiter and
Earth (where $\langle\rho\rangle$ $\sim 1.3$ g cm$^{-3}$ and $\sim5.5$
g cm$^{-3}$, respectively). As a result, an Earth-like planet can be
tidally disrupted by a giant planet like our Jupiter, but very well
may survive if the radius of the Jovian planet is somewhat larger (so
that its density is lower). Since both Hot Jupiters and young Jupiters
(in any orbit) have inflated radii, incoming rocky planets are likely
to survive tidal disruption, reach their centers, and increase the
masses of their cores.

\begin{figure} 
\figurenum{10} 
{\centerline{\epsscale{0.90} \plotone{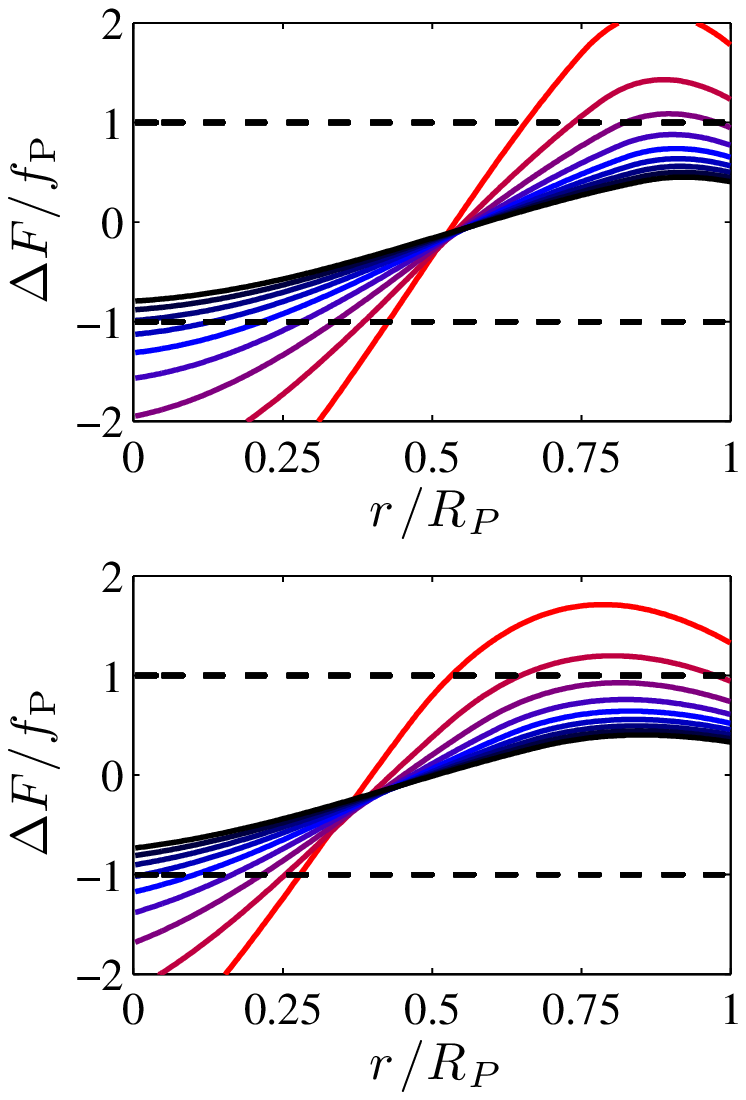} } } 
\figcaption{Ratio of tidal forces to rocky planet surface gravity 
$\fgp$ as a function of radial distance (where $\fgp$ = 
$G\mrock/\rrock^2$).  The Jovian planet is taken to have no core, mass
$M_P=M_J$, radius $R_P=R_J$, and $n=1$.  Results are shown for rocky
planets with mass $\mrock$ = 1 $\mearth$ (top panel) and $\mrock$ = 10
$\mearth$ (bottom panel).  In each panel, the curves correspond to
different mean densities of the rocky planet, ranging from $\rhorock$
= 2 g cm$^{-3}$ (top red curve) to $\rhorock$ = 10 g cm$^{-3}$ (bottom
black curve).  The region between the dashed lines indicates radial
locations where the rocky planet can survive. }
\label{fig:tides1} 
\end{figure}

\begin{figure} 
\figurenum{11} 
{\centerline{\epsscale{0.90} \plotone{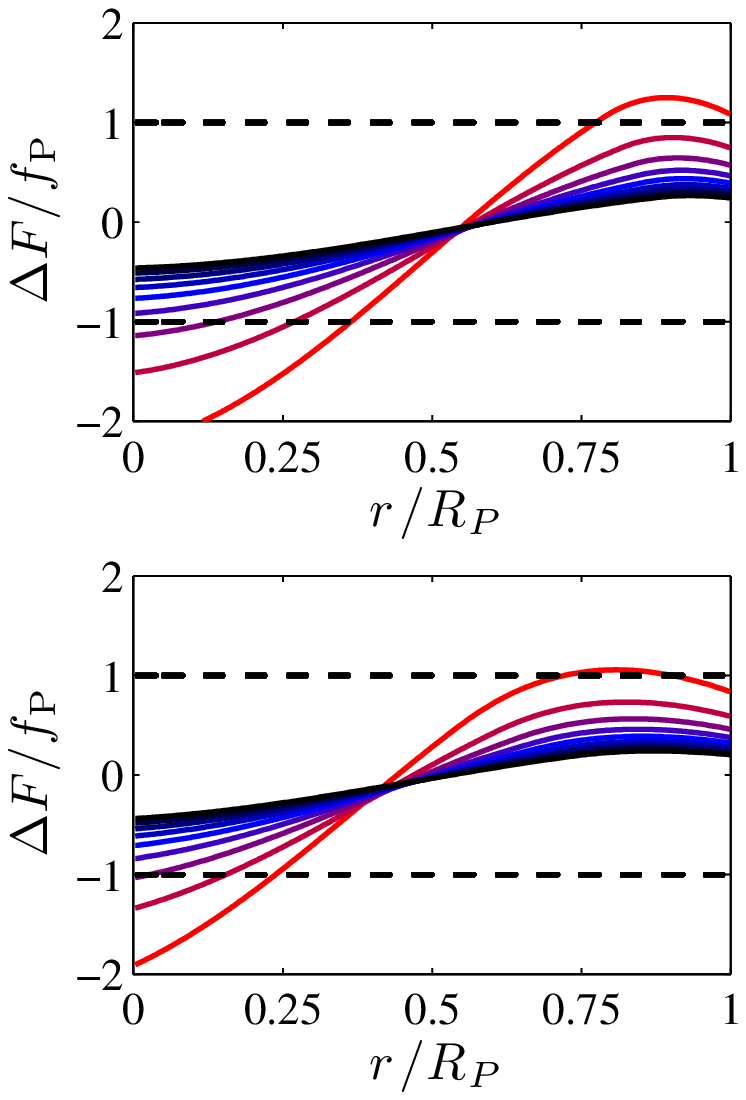} } } 
\figcaption{Ratio of tidal forces to rocky planet surface gravity as 
a function of radial distance for an inflated gaseous planet. Here 
the Jovian planet is taken to have no core, mass $M_P=M_J$, radius 
$R_P=1.2R_J$, and $n=1$.  Results are shown for rocky planets with
mass $\mrock$ = 1 $\mearth$ (top panel) and $\mrock$ = 10 $\mearth$
(bottom panel).  In each panel, the curves correspond to different
mean densities of the rocky planet, ranging from $\rhorock$ = 2 g
cm$^{-3}$ (top red curve) to $\rhorock$ = 10 g cm$^{-3}$ (bottom black
curve).  The region between the dashed lines indicates radial
locations where the rocky planet can survive. }
\label{fig:tides2} 
\end{figure}

The ratio of the tidal force to the surface gravity of the rocky
planet can be calculated as a function of radial distance using
equation (\ref{tidalexact}). Here we explore a variety of planetary
properties.  The Jovian planet mass was held constant at $M_P=1M_{J}$
and the radius varied over the range $R_P=1-2R_J$, leading to average
densities $\langle\rho\rangle\sim 0.1-1.0$ g cm$^{-3}$.  The results
are shown in Figure \ref{fig:tides1} for $R_P=1R_J$ and in Figure
\ref{fig:tides2} for $R_P=1.2R_J$, where each curve represents a
different rocky planet density. The latter value corresponds to a
typical radius for observed Hot Jupiters, which are inflated relative
to expectations \citep[e.g.,][]{laughlin}. For even larger values of
$R_P$ (not shown), the densities of the Jovian planet are small enough
that little tidal disruption takes place.  The expected range of
(mean) densities for rocky planets is $\rho\approx2-10$ g cm$^{-3}$
\citep{valencia}, with planet masses $\mrock$ = 0.1 -- 20 $\mearth$.
Note that planets less massive than $\sim1\mearth$ are not expected
to survive. Most of the calculations were performed for coreless
polytropes of index $n=1$; adding a core does not significantly alter
the strength of the tidal forces (and hence these curves).

Notice that the tidal force $\Delta F$ vanishes at radii
$r\sim{R_P}/2$. The tidal force becomes negative for smaller radii,
indicating that an orbiting planet will be first be stretched and
subsequently compressed during its inward trajectory. In the center of
the Jovian planet (where $r\to0$), the compression reaches a maximum
and the force given by equation (\ref{usualtide}) formally approaches 
a well-defined limit 
\be 
\lim_{r\to 0}\Delta F = - {8\pi\over3} G \rho_c \, \rrock \, . 
\ee
Note that this is a (mathematically) formal result, because we are
taking the limit $r\to0$ but the rocky planet has a finite size
$\rrock$.

The area between the dashed lines (set where $(\Delta{F})/\fgp=\pm1$)
in Figures \ref{fig:tides1} and \ref{fig:tides2} indicates the regions
where the surface gravity of the rocky planet can withstand the tidal
acceleration.  If the ratio $(\Delta{F})/\fgp>1$, the outer layers of
the rocky planet are stripped away; if $(\Delta{F})/\fgp<-1$, the
planet experiences devastating levels of compression. Here we consider
a rocky planet to survive whenever the condition $|(\Delta F)/\fgp|<1$
is satisfied all the way to the core.  All but the smallest planets
are able to penetrate deep into the interior (at least $50-70\%$ of
the distance to the center) before experiencing severe tidal
disruption; in many cases, the rocky planet reaches the core intact.
As expected, the small planets ($\mrock\sim0.1\mearth$) must be very
dense to survive the entire trajectory, and we exclude them in the
remaining discussion. Here we focus on rocky planets with masses
$\mrock$ = 1, 10, and 20 $\mearth$.  For Jovian planets with the mean
density of Jupiter (1.2 g cm$^{-3}$), a rocky planet with $\mrock$ =
$1\mearth$ must have density greater than $\rhorock\approx5-6$ g
cm$^{-3}$ for survival; larger rocky planets with $\mrock = 10\mearth$
can survive with somewhat lower density $\rhorock\approx3-4$ g
cm$^{-3}$.  On the other hand, for Jovian planets with sufficiently
high densities (roughly, where the inequality of equation
[\ref{tidalapprox}] is violated), the tidal forces at the surface can
exceed the rocky planet surface gravity, and such planets can be
tidally disrupted before even entering the atmosphere.

After a rocky planet is tidally disrupted, its fate remain unclear.
The resulting rocky debris is unlikely to continue the journey toward
the core, because smaller objects are more susceptible to frictional
dissipation (and further tidal disruption). The rocky debris is likely
to remain in the outer envelope of the giant planet and enrich the
atmosphere, but its longer-term fate should be the subject of further
study. The outcome will depend on, among other things, the opacity of
the Jovian planet. For example, high opacities lead to convection,
which could cause the rocky debris to be uniformly mixed into the
gaseous planetary interior.

In summary, we find that large rocky planets ($\mrock\sim10\mearth$)
have a good chance of reaching the core intact; even planets with
$\mrock\sim1\mearth$ are able to survive in many cases, as long as the
Jovian planet density is sufficiently low. As a result, these
collisions provide a viable mechanism for allowing significant amounts
of rocky material to be be accreted onto the cores of gaseous planets,
especially those with inflated radii like young Jovian planets and
much of the observed Hot Jupiter sample. 

\section{Conclusion}
\label{sec:conclude} 

This paper explores collisions between Jovian planets and smaller
terrestrial bodies in order to identify possible changes in the
structure of the larger planets. This work focuses on the scenario
where the Jovian planet has migrated inward and entered into a tight
orbit, and subsequently experiences collisions with rocky planets that
migrate later. However, many of the results apply to other parts of
parameter space, e.g., the possible accretion of cores by gaseous
planets formed via gravitational instability
\citep[e.g.,][]{boss,boley,boley2,helled}.  These collisions can
affect the structure of the target giant planets in three ways: [a]
The metallicity is increased, [b] The core size can be increased, and
[c] A substantial amount of energy is imparted to the giant
planet. For Hot Jupiters, these impacts could thus help explain the
observed anomalous radius distribution.

Rocky planets are efficiently captured by giant planets and thereby
increase the metallicity of the larger body. For sufficiently high
impact speeds $v_0$ and large impact parameters $x_0>x_\ast$, the
projectile planet can pass through the giant planet and emerge out the
other side. For nearly head-on collisions, the initial velocity
required to to escape capture is much higher ($\vpen \sim 10^4$ km
s$^{-1}$) than the expected impact speeds $v_0\approx40-150$ km
s$^{-1}$ \citep{ketchum}.  With these latter speeds, incoming planets
can escape only for large impact parameters $x_0 \gta 0.8R_P$, where
the densities are low. In other words, penetration is possible when
the incoming rocky planet skims the surface of the Jovian target.  As
a result, most rocky planets lose enough kinetic energy through their
initial passage that they remain gravitationally bound and are
captured (see Figure \ref{fig:vpen}; an analytic estimate of the
required penetration speeds is given in Appendix
\ref{sec:speedappendix}). Nonetheless, the capture cross section of
the giant planet can be reduced; for typical collision speeds
$v_0\sim100$ km s$^{-1}$, the cross section is smaller than the
geometrical area by $\sim30\%$.

These collisions provide a plausible mechanism for explaining the
massive cores that have been inferred for observed Hot Jupiters.
Although metallicity increases are essentially automatic through such
collisions, core masses can only increase if the rocky planets avoid
tidal disruption and survive to reach the central regions. The
expected parameter space includes cases where the impinging planets
survive and cases where they are completely disrupted. As a benchmark,
rocky planets with the mass and radius of Earth live near the
threshold for tidal destruction when they collide with giant planets
having the mass and radius of Jupiter.  However, sufficiently dense
projectile planets or inflated giant (target) planets allow for the
rocky planets to survive all the way to the central regions (see
Figures \ref{fig:tides1} and \ref{fig:tides2}).  This finding is
relevant because such collisions are expected to take place early in
the system's evolution when the giant planets are larger, and also
because Hot Jupiters are subject to additional heating sources and are
thus larger in radius for a given mass. As noted above, this process
also provides a mechanism through which planets formed via
gravitational instability can acquire cores.

For collisions to significantly alter the metallicity or the core mass
of a giant planet, it must accrete rocky bodies that represent total
masses $\Delta{M}\sim10-100\mearth$. The accumulation of such mass
increments could take place through collisions with many smaller rocky
bodies or via fewer larger bodies. Large rocky bodies, with mass
$\mrock>10\mearth$, are more effective for several reasons: First,
larger rocky bodies are subject to less eccentricity damping during
migration, and are thus more likely to collide with the giant planets.
More specifically, eccentricity damping will be small enough to allow
collisions when the rocky planets are large enough to partially clear
gaps \citep{arty,kley}, which occurs when their Hill sphere exceeds
the disk scale height \citep{crida,migration}. These considerations
indicate that relatively large rocky planets (with mass $\mrock$
$\approx10-20\mearth$) are favored in order to achieve partial
gap-clearing, reduced eccentricity damping, and ultimately collisions.
In addition, collisions with larger rocky bodies allow for the
required mass increments to be realized through a lower number of
encounters. Finally, larger rocky bodies are more likely to survive
tidal disruption and reach the central regions (see Figures
\ref{fig:tides1} and \ref{fig:tides2}). On the other hand, if the
circumstellar disk has high levels of turbulence, small rocky planets
are subject to stochastic migration \citep[e.g.,][]{adamsbloch}, and
can still experience encounters with gas giants. Such encounters often
result in collisions \citep{ketchum} when they take place at small
semimajor axes ($a\lta0.1$ AU); for larger $a\gta1$ AU, most
encounters result in ejection of the smaller rocky planet, with
collisions taking place only $\sim10\%$ of the time
\citep{ketchumalt}.

Collisions provide a substantial amount of energy to the target giant
planet. For rocky planets with mass $\mrock=10\mearth$ and initial
speed $v_0=60$ km s$^{-1}$, the kinetic energy transferred to the
giant planet is $\Delta{E}\sim10^{42}$ erg.  If, for example, this
energy is radiated over a time scale of 1 Gyr, the associated
increment in planetary luminosity would be $\Delta{L_P}\approx4$
$\times10^{18}$ W, which is large enough to affect the internal
structure of the planet \citep{bodenheimer}.  However, in order for
these collisions to provide an effective (long-term) heating source,
most of the energy must be deposited deep in the Hot Jupiter interior,
in regions of high opacity.  For a given rocky planet mass, the amount
of energy deposited depends on the density of the Jovian planet.  For
a Jovian planet with the mass and radius of Jupiter, most of the
kinetic energy is dissipated in the outer layers of the planet. At the
other end of the possible range, for a Jovian planet with the mass of
Jupiter and an inflated radius $R_P=2R_J$, little energy is dissipated
in the outer layers and a sizable fraction remains upon impact with
the core (see Figures \ref{fig:energyv0} and \ref{fig:energyx0}, and
Appendix \ref{sec:energyappendix}). 

An important unresolved issue is the question of how quickly the
collision-induced energy is transferred out of the planet. Collisions
with large terrestrial planets deposit enough energy in the central
regions of the Jovian planets to potentially affect their structure.
This energy could even play a role in explaining the radius anomalies
of the transiting exoplanets if the time scales for energy loss are
sufficiently long. In future work, interior models of Hot Jupiters
should be modified to include this possible heat source.

As shown herein, incoming rocky bodies are sometimes tidally destroyed
after entering the surface of the giant planet (see also \cite{anic}
and \cite{li} for detailed simulations of specific cases). This work
indicates that most planets heavier than $\mrock\sim10\mearth$ are
able to withstand tidal disruption (see Figures \ref{fig:tides1} and
\ref{fig:tides2}). However, this work focuses on frictional forces and
tidal disruption, but does not consider the possible destruction of
incoming rocky planets through heating and ablation.  Through both
ablation and tides incoming planets continuously lose mass throughout
their descent, and the relative importance these effects should be
studied in future work. If ablation is strong enough, the rocky
planets could completely disintegrate before reaching the core, even
if they can survive tidal disruption. Ablation will thus increase the
overall metallicity of the Jovian planet atmosphere and reduce the
amount of kinetic energy and mass delivered to the core.  Since the
calculations of this paper do not include ablation, the quantities of
heat and heavy elements deposited deep in the Jovian planet interior
could be lower than estimated herein.

Although ablation effects are important for smaller rocky bodies such
as planetesimals \citep{benbru}, they are not expected to play a
dominant role for full-sized planets with masses $\mrock$ = 1 -- 10
$\mearth$ \citep{anic,li}.  In rough terms, an incoming projectile
planet is expected to be destroyed only after it encounters a gas mass
larger than its own mass \citep{koryzan}. The mass in gas $\Delta m$
encountered on an inward trajectory is roughly given by $\Delta m$ =
$\cross \rho_P R_P$, so that survival requires a minimum size
$\rrock\gta3\rho_P R_P/(4\rhorock) \sim 1 - 2 \rearth$. As a result,
the larger rocky planets of interest here are likely to survive
ablation, but smaller bodies are susceptible to destruction.

The results of this paper show that the mean density of the giant
planet plays a defining role in determining the consequences of
these collisions. The average density helps determine whether or not
incoming rocky planets can survive tidal disruption and affects the
radial layers where kinetic energy is deposited. Since giant planets
are born with enlarged radii, this density dependence translates into
a dependence on the age of the planet.  A rocky planet that collides
with a young giant planet (with low density) can have dramatically
different consequences than a rocky planet that collides with a mature
giant planet (with higher density).  The environment in which the
giant planet resides plays an additional role. Since Hot Jupiters
often remain significantly inflated throughout their evolution, this
subset of extrasolar planets should be even more sensitive to these
collisions. This lends promising support for the proposal that
planetary impacts could help explain, in part, the anomalous radius
distribution of the observed Hot Jupiters.

\acknowledgments

This paper resulted from the senior thesis of KA and benefited from
discussions with many colleagues, including Tony Bloch, Jake Ketchum,
and Greg Laughlin. This work was supported by NASA grant NNX11AK87G
from the Origins of Solar Systems Program, and by NSF grant
DMS-0806756 from the Division of Applied Mathematics.

\newpage

\appendix 

\section{Analytic Estimate for the Penetration Velocity} 
\label{sec:speedappendix} 

In this Appendix, we derive an analytic estimate for the penetration
velocity $\vpen$, the speed required for an impinging rocky planet to
pass through a Hot Jupiter and thereby avoid capture.  As shown below,
the required velocity is much larger than the gravitational escape
speed $\vesc$ from the planet. As a result, gravity can be neglected
to leading order, and the equation of motion simplifies to the form 
\be 
\frac{d \bf{v}}{d \tau} = -\alpha f^n v^2 \vhat \nonumber = 
-\alpha f^n v (\dot{x} \xhat + \dot{y} \yhat) \, ,
\label{friction}
\ee 
where all of the variables are dimensionless (defined in the text). By
definition, the initial velocity lies in the $\yhat$ direction and the
initial spatial position is given by $y_0 = - L$, where $L>0$.
Further, since the required speeds are large, the path of the rocky
planet through the Hot Jupiter can be approximated as a straight line
(for cases of interest where the rocky planet is not captured). As a
result, we can set $\dot{x}\approx{0}$ and $v=\dot{y}$.  With these
simplifications, equation (\ref{friction}) becomes
\be
\frac{dv}{d \tau} = -\beta v^2, \qquad \beta \equiv 
\alpha \langle f^n \rangle \, , 
\label{dvdtau} 
\ee
where we have made the approximation $f^n=\langle{f^n}\rangle$. This
differential equation is easily integrated.  Applying the boundary
condition $v=v_0$ at time $\tau$ = 0, we obtain the result 
\be
v = \frac{dy}{d \tau} = \frac{v_0}{1 + \beta \tau v_0}.
\label{dydtau}
\ee
Integrating equation (\ref{dydtau}) for $y(\tau)$ yields
the solution 
\be
y - y_0 = \beta^{-1} \ln(1 + v_0 \beta \tau) \, . 
\label{ysolution} 
\ee
We can combine equation(\ref{dydtau}) and (\ref{ysolution}) 
to find the speed $v$ as a function of position $y$, 
\be
v = v_0 \, {\rm e}^{-\beta(y - y_0)}.
\label{escape}
\ee
In order for the rocky planet to penetrate the Jovian planet, the
speed must exceed the escape speed ($v > \vesc$) at the location
$y=+L=-y_0$. The minimum initial speed $\vpen$ required for
penetration is thus given by the condition
\be
\vesc = \vpen \, {\rm e}^{-2 \beta |y_0|}, \qquad {\rm or} 
\qquad \vpen = \vesc \, {\rm e}^{2 \beta |y_0|} \, .
\label{penetrate}
\ee 
Note that the escape speed is given by $\vesc=(2\mu_0/\xi_0)^{1/2}$
in terms of dimensionless variables. 

Next we find the dependence of the penetration speed $\vpen$ on the
impact parameter $x_0$. Here we estimate the average density by
approximating the density profile of the Jovian planet as a Gaussian
function, where the peak is a function of the impact parameter.  For
head on collisions, with $x_0=0$, we assume $\langle f^n \rangle$ 
$\approx 1/2$.  The constant $\beta$ is thus given by 
\be
\beta = \frac{1}{2}\alpha 
\exp\left[- \gamma \, x_0^2 \right]\,,
\label{beta}
\ee
where we have introduced an unspecified parameter $\gamma$, which is
expected to be of order unity. We find that reasonable agreement with
numerical results can be obtained using $\gamma\approx1/4$ (see Figure
\ref{fig:vpen}).  Equations (\ref{penetrate}) and (\ref{beta})
together yield a simple relation between the initial velocity and the
impact parameter.  In the limiting case of grazing collisions, where
$x_0=\xi_0$ and $y_0=0$, the penetration speed is equal to the escape
velocity $\vpen=\vesc$.  On the other hand, for head-on collisions
where $x_0=0$ and $y_0=-\xi_0$, the penetration speed attains its
maximum value $\vpen\approx\vesc\exp[\alpha\xi_0]$. Note that even 
though the parameters $\alpha$ and $\xi_0$ are ``of order unity'',
their typical values are $\sim3$, so that the exponential factor is
large and $\vpen\gg\vesc$.

The above derivation assumes that $\langle{f^n}\rangle\approx1/2$.
For the particular case of an $n=1$ polytrope with no central core
($\xi_c\to0$), we can directly test this approximation:
\be
\langle{f^n}\rangle = \langle{f}\rangle = 
{1 \over \pi} \int_0^\pi {\sin\xi\over\xi}d\xi = 
{1 \over \pi} \sineint(\pi) \, , 
\ee
where $\sineint(x)$ is the sine integral \citep{handbook}, which 
has the known value $\sineint(\pi) \approx 1.85194$, so that 
$\langle f^n \rangle \approx$ 0.5895. Thus, the approximation is 
reasonable. 

\section{Analytic Estimate for Energy Dissipation} 
\label{sec:energyappendix} 

In this Appendix, we provide an analytic estimate for the energy $E_f$
remaining when an incoming rocky planet reaches the core of the Jovian
planet. The incoming planet has initial kinetic energy $E_0=v_0^2/2$,
where the speed $v_0$ is evaluated when the rocky planet reaches the
surface of the Jovian planet.  For the particular case of direct
collisions, with zero impact parameter, the incoming planets travel
along a radial path from the gas giant surface to the core. The
amount of energy $W$ dissipated along this path is given by 
\be
W = \alpha \int_{\xi_c}^{\xi_0} \!  f^n v^2  \, d\xi \,,
\ee
where we have chosen signs so that $W>0$, and where all quantities are
dimensionless (see text).  Using equation (\ref{escape}) to estimate
the speed as a function of position, the work $W$ can be expressed as 
\be
W = \alpha {v_0}^2 \int_{\xi_c}^{\xi_0} \! f^n 
{\rm e}^{2 \beta(\xi - \xi_0)}  \, d\xi \, , 
\ee
where $\beta = \alpha \langle f^n\rangle$.  If we make the simplifying
approximation $f^n \approx 1/2$, the expression becomes 
\be
W = {1 \over 2} \alpha {v_0}^2 \int_{\xi_c}^{\xi_0} \!  
{\rm e}^{\alpha(\xi - \xi_0)} \, d\xi,
\ee
which can be integrated to yield
\be
W = {1 \over 2} {v_0}^2
\left[ 1 - {\rm e}^{\alpha(\xi_c - \xi_0)} \right] = 
E_0 \left[ 1 - {\rm e}^{\alpha(\xi_c - \xi_0)} \right]\,. 
\ee
Conservation of energy implies $E_f$ = $E_0 - W$, so the remaining
energy when these planets reach the core is given by 
\be
E_f = {\rm e}^{\alpha(\xi_c - \xi_0)} E_0 \, . 
\label{efinal} 
\ee
Consider a giant planet with polytropic index $n=1$ and core 
radius $\xi_c=1$. The final energies $E_f$ predicted by equation
(\ref{efinal}) for incoming rocky planets with masses $\mrock$ = 1,
10, and 20 $\mearth$ are given by $E_f /E_0 \sim 0.03$, 0.20, and
0.30, in rough agreement with the numerically determined results.


\end{document}